\documentclass[aps,prd,superscriptaddress, twocolumn]{revtex4}%
\usepackage{amsfonts}
\usepackage{txfonts}
\usepackage{amssymb}
\usepackage{amsbsy} 
\usepackage{epsfig}

\def\be{\begin{equation}} \def\ee{\end{equation}}
\def\bea{\begin{eqnarray}} \def\eea{\end{eqnarray}}

\def\nn{\nonumber}

\def\bsigma{{\boldsymbol \sigma}}

\def\balpha{{\boldsymbol \alpha}}
\def\bpartial{{\boldsymbol \partial}}
\def\bq{{\bf q}}

\def\bx{{\bf x}}

\def\bp{{\bf p}}
\def\bE{{\bf E}}
\def\bB{{\bf B}}
\def\bD{{\bf D}}

\def\bA{{\bf A}}
\def\bG{{\bf G}}
\def\ba{{\bf a}}
\def\bb{{\bf b}}
\def\bj{{\bf j}}
\def\by{{\bf y}}
\def\ba{{\bf a}}
\def\be{{\bf e}}
\def\bz{{\bf z}}
\def\bv{{\bf v}}
\def\bX{{\bf X}}
\def\bY{{\bf Y}}

\def\la{\langle}
\def\ra{\rangle}

\def\rw{\rightarrow}

\begin{document}

\title{ On the Interaction of Electrons, Magnetic Monopoles, and Photons }

\author{Zhong Wang}
\affiliation{ Institute for
Advanced Study, Tsinghua University, Beijing,  China, 100084}

\affiliation{Collaborative Innovation Center of Quantum Matter, Beijing 100871, China }


\date{ \today}

\begin{abstract}

We study quantum systems of interacting electrons, magnetic monopoles, and electromagnetic field. We formulate a convenient field theory, in which the electron-photon, monopole-photon, and electron-monopole interactions take simple forms.

\end{abstract}

\pacs{14.80.Hv, 11.10.Ef, 12.20.Ds, 11.15-q}

\maketitle

\section{Introduction}

As is well-known from the early times, inclusion of both electric and magnetic charges restores the symmetry between electricity and magnetism in the Maxwell equations\cite{jackson1998}. Several years after the birth of quantum mechanics, Dirac pointed out\cite{dirac1931} that the magnetic monopole, if it exists, cannot carry an arbitrary amount of magnetic charges. In fact, the minimal magnetic charge $g$ and the minimal electrical charge $e$ must satisfy a quantum condition, $eg=2\pi n$ ($n$ is an integer), which is known as the Dirac quantization condition. The magnetic monopole has generated enduring interest in many fields of physics\cite{shnir2005,milton2006,hooft1974,guth1980}.
For instance, it has played a crucial role in theoretical developments related to electromagnetic duality\cite{montonen1977,seiberg1994,seiberg1994a,harvey1997}.

In Dirac's original theory of monopoles, he used the auxiliary concept of ``Dirac string''.
In a work\cite{dirac1948} on classical action in 1948, Dirac introduced dynamic variables for the string. As a result, the entire theory became very complicated.  A much more transparent formulation was put forward by Wu and Yang\cite{wu1975,wu1976a}, who borrowed the idea of fiber bundle from mathematics. In this language, systems with monopoles are identified with nontrivial $U(1)$ bundles. To describe them quantitatively, overlapping patches of coordinates were used, and various quantities ( such as gauge potential ) follow prescribed transformation rules upon changing from one patch to another. In this formulation the troublesome Dirac string is absent.

Based on the language of fiber bundle, an elegant classical
Lagrangian for monopoles was proposed in Ref.\cite{wu1976}. A
notable feature in this theory is that the action becomes
multi-valued. The natural next step is to develop this theory
into a form that describes quantum systems of interacting electrons,
monopoles, and photons, however, this is not a straightforward
problem \footnote{In C N Yang's \emph{Selected papers
(1945-1980)}\cite{yang2005}, there is his following commentary to
Ref.\cite{wu1976}: ``...and we started to develop a second quantized
theory using path integral. This effort was, however, frustrated.''}.
Without invoking this Lagrangian formulation, in a prescient
work Tu, Wu, and Yang developed a
Hamiltonian formalism\cite{tu1978}, in which inspired guesswork was needed. Several other efforts on the quantum field theory of electron-monopole-photon systems can be found in Ref.\cite{blagojevic1985} and the references therein.

Aiming at a more convenient formulation for interacting electrons, monopoles, and photons,
in the present paper we reinvestigate the classical Lagrangian of
Wu and Yang\cite{wu1976}. With this classical Lagrangian as a hint, we formulate a simple quantum field theory for electron-monopole-photon systems. This is done in the path integral approach.
Simple forms for electron-photon, monopole-photon, and
electron-monopole interactions are obtained in a natural manner.

The remainder of this paper is organized as follows. In
Sec.\ref{sec:m-photon}, we study the interaction of monopoles and
photons (but no electron). We work on this case first because it
illustrates the way monopole-photon interaction emerges in this
approach. We also find in our formalism that a fermionic Dirac
monopole has an intrinsic electric dipole moment with a $g$-factor
$2$. In Sec.\ref{sec:e-m-photon} we proceed to formulate the
interactions of electrons, magnetic monopoles, and photons. The
complete Lagrangian is also contained in Sec.\ref{sec:e-m-photon}. In
Sec.\ref{sec:motion} we shall study both classical and quantum
equations of motion, followed by Sec.\ref{sec:dual}, in which we
present a dual description.  In Sec.\ref{sec:consistent} we shall
study the effective monopole-monopole interaction in both the
original and the dual descriptions, which can be regarded as a
consistency check of our formulation.

\section{Interaction of magnetic monopoles and photons}\label{sec:m-photon}

In this section we formulate a field theory for interacting magnetic monopoles and photons (without electrons).  The magnetic charge is treated as a topological charge; thus, it is unnecessary to introduce a minimal coupling for monopoles. The basic idea dates back to Dirac\cite{dirac1948}  (see also Wu and Yang\cite{wu1976}), however, therein the monopole was treated as a classical particle instead of a quantum field, furthermore, the interaction between monopole and electromagnetic field was implicit in the Lagrangian. Later developments of quantum field theory of magnetic monopoles\cite{blagojevic1985} were often complicated by the Dirac string.  In the present approach, the Dirac string is absent because of a convenient separation of electromagnetic field into two parts, the first part being dynamics, and the second part being kinematic.

Let us start from the Maxwell equations \cite{jackson1998}
\bea \nabla\cdot{\bf B} = g\rho_m , &&\,\,\,\,\, (a) \nn \\
\nabla\times{\bf E} + \frac{\partial \bB}{\partial t} = -g{\bf j}_m , &&\,\,\,\,\, (b) \nn \\
\nabla\cdot{\bf E} = e\rho_e , &&\,\,\,\,\, (c) \nn \\
\nabla\times{\bf B} - \frac{\partial \bE}{\partial t} =e{\bf j}_e , &&\,\,\,\,\, (d)  \nn \label{maxwell}
\eea  where $(\rho_e,\bj_e)$ are the density and currents of electrons \footnote{In this paper ``electron'' refers to any particle with electric charge $e$ (we take the convention that $e>0$). It can be regarded as the positron or the proton. }, $(\rho_m,\bj_m)$ are the density and currents of magnetic monopoles, and $\bE,\bB$ are electric and magnetic fields. In the absence of electrons, we can just take $\rho_e=\bj_e=0$. We would like to emphasize that, in classical physics\cite{wu1976}, two of the Maxwell equations [(a) and (b)] should be regarded as constraints (kinematic equations), which are not derived from variation of action.  The other two Maxwell equations, (c) and (d), are derived from the action principle  \footnote{ Due to electromagnetic duality, we can also regard the last two equations as constraints, while the first two as derived from variational principle. This is the dual description.}.
In the quantum mechanical formulation, we regard $(a)$ and $(b)$ as kinematic equations and exploit their consequences. Let us begin with the first equation, namely, $\nabla\cdot\bB =g\rho_m$. We separate out a part of $\bB$, such that the rest has vanishing divergence. In mathematical form, we have
\bea \bB=\bb+ \bb' \label{h-decomposition} \eea in which  \bea \bb'(\bx,t) = g\int d\by \bG(\bx,\by)\rho_m (\by,t),  \label{hprime} \eea  with the shorthand notation \bea \bG(\bx,\by)=-\nabla \frac{1}{4\pi |\bx-\by|} =\frac{\bx-\by}{4\pi |\bx-\by|^3}  \eea  It is readily seen that $\bb'(\bx,t)$ satisfies $\nabla\cdot\bb'(\bx,t)=g\rho_m(\bx,t)$, therefore, we have $\nabla\cdot\bb(\bx,t) =0$, as we required.  It is therefore possible to introduce a vector potential $\ba$ such that \bea \bb=\nabla\times\ba \label{h-a} \eea

Next we would like to do a similar trick for the electric field. This
step is less straightforward. Here, we make use of the second
kinematic equation, namely $(b)$.  We split the electric field as
$\bE=\be +\be'$, such that the kinematic part $\be'$ satisfies  \bea
\nabla\times\be' +\partial_t\bb'= -g\bj_m \label{eprime} \eea
Together with Maxwell equation $(b)$, it also implies  \bea
\nabla\times\be +\partial_t\bb= 0 \label{eh} \eea Since the explicit
formula for $\bb'$ has already been given, we can use
Eq.(\ref{eprime}) to determine $\be'$, for which we indeed find a
solution    \bea \be'(\bx,t)=g\int d\by \bG(\bx,\by)\times
\bj_m(\by,t) \label{solution-eprime} \eea  To avoid distractions by
too many details, we have left the calculations for this to Appendix
\ref{sec:derivation1}.

One of the merits of splitting $\bE$ as $\be+\be'$ is as follows.
Inserting Eq.(\ref{h-a}) into  Eq.(\ref{eh}), we have $\nabla\times(\be+\partial_t\ba)=0$, which allows us to write down $\be
+ \partial_t\ba = -\nabla a_0$, or equivalently, $\be = -
\partial_t\ba -\nabla a_0$.  Summarizing these equations, we have
the following mathematical expressions for the electric and magnetic
fields in terms of gauge potential \bea \bE= -\partial_t\ba-\nabla
a_0 +\be'; \,\,\, \bB =\nabla\times\ba +\bb' \label{EH} \eea If the
magnetic monopole is absent, then $\bb'=\be'=0$, and these formulas
are reduced to the usual ones. Roughly speaking, in Eq.(\ref{EH}) we have separated the electromagnetic field into two parts, the dynamic part described by $\be$ and $\bb$, and the kinematic part described by $\be'$ and $\bb'$. The latter is fully determined by the monopole density and current. It is worth emphasizing that the two equations obtained in Eq.(\ref{EH}) are natural consequences of $(a)$ and $(b)$ of the Maxwell equations.

Now that we have written $\bE$ and $\bB$ in convenient forms,
we are ready to present the path integral formulation. The Lagrangian density of the monopole-photon system is given as \bea L= L_f + L_g
+ L_\lambda \eea where the first part is \bea L_f =\frac{1}{2}(\bE^2-\bB^2)\, , \eea $\bE$ and $\bB$ being given by Eq.(\ref{EH}).
The second part of $L$, namely $L_g$, is the Lagrangian density of monopoles. If we take the monopoles as
Dirac particles (though our results are not limited to Dirac
particles), we have \bea L_g = \psi^\dag(i\partial_t  + i\sum_{i=1}^3
\alpha_i\partial_i -\beta M)\psi \eea where $\alpha_i$ and $\beta$
are Dirac matrices, and $M$ is the (bare) mass of monopoles. The
monopole density is $\rho_m = \psi^\dag\psi$, and the monopole current
is $\bj_m= \psi^\dag\balpha\psi$ (or
$j_m^i= \psi^\dag\alpha_i\psi$).

As the last part of the Lagrangian density, $L_\lambda$ is a gauge fixing term\footnote{Different gauge fixing terms amount to different gauge choices (in the Faddeev-Popov approach\cite{faddeev1967}). See Ref.\cite{peskin1995} or Ref.\cite{huang1982} for more details. }, which is added to ensure that the photon
propagator is nonsingular. The nature of this term in our formulation is the same as that in usual quantum electrodynamics without monopole\cite{peskin1995}, therefore, we shall not discuss it in more detail. For simplicity, we take $L_\lambda$ to be \bea L_\lambda =
\frac{1}{2\lambda} (\nabla\cdot\ba)^2 \eea though other choices are also
allowed.

Now the full quantum theory is based on the path
integral\footnote{Here $\psi^\dag,\psi$ are regarded as Grassmann
numbers when monopoles are fermionic. Our formulation is also valid
when monopoles are bosons, in which case $\psi^\dag,\psi$ are
replaced by ordinary number. Similarly for electrons. }
\bea \int  D\ba Da_0 D\psi^\dag D\psi\exp[i\int dt d^3x L(\bx,t)] \eea

To achieve a convenient field theory of interacting magnetic
monopoles and photons, we have to investigate the Lagrangian density
in more depth. A peculiar feature is worth noting.  At first sight,
it seems that the partial derivative $\partial_i$ in $L_g$ should be
replaced by a covariant derivative,  describing the monopole-photon
interaction. Somewhat surprisingly, this is unnecessary and the
partial derivative suffices.  In fact, the monopole-photon
interaction has already been included in $L_f$. To see this fact, we can expand $L_f$ as \bea L_f &=& \frac{1}{2}(\bE^2-\bB^2) \nn \\
&=& \frac{1}{2}[(-\partial_t\ba-\nabla
a_0+\be')^2-(\nabla\times\ba+\bb')^2]    \nn \\ &=&
\frac{1}{2}[(-\partial_t\ba-\nabla a_0)^2 - (\nabla\times\ba)^2
+(\be')^2-(\bb')^2] \nn \\ && -(\partial_t\ba+\nabla a_0)\cdot\be'
-(\nabla\times\ba)\cdot\bb'  \eea After integration by parts and
discarding total derivatives\footnote{For instance, the last term can
be simplified as
$-(\nabla\times\ba)\cdot\bb'=\nabla\cdot(\bb'\times\ba) -
\ba\cdot(\nabla\times\bb')=\nabla\cdot(\bb'\times\ba)$, which is a
total derivative. }, we have \bea L_f
&=& \frac{1}{2}[(-\partial_t\ba-\nabla a_0)^2 - (\nabla\times\ba)^2
+(\be')^2-(\bb')^2] \nn \\ && -(\partial_t\ba )\cdot\be' \label{L-f-simp-2} \eea
Now the complete Lagrangian density of monopole-photon system is \bea && L_f +L_\lambda +L_g \nn \\
&=&\frac{1}{2} (-\partial_t\ba-\nabla a_0)^2 - \frac{1}{2}(\nabla\times\ba)^2  +\frac{1}{2\lambda}(\nabla\cdot\ba)^2  \nn \\ && +   \psi^\dag(i\partial_t  + i\sum_{i=1}^3
\alpha_i\partial_i -\beta M)\psi \nn \\ && +\frac{1}{2}
(\be')^2-\frac{1}{2}(\bb')^2  -(\partial_t\ba )\cdot\be' \label{L-f-simp} \eea
Let us figure out the physical meanings of these terms. The first three
terms of Eq.(\ref{L-f-simp}) give
rise to the familiar bare photon propagator (propagator of $(a_0,\ba)$), which
can be found in textbooks. We list them as\cite{huang1982} \bea D^{ij}(q) &=& -\frac{1}{q^2+i\epsilon}(\delta_{ij}-\frac{q_iq_j}{\bq^2})+\frac{\lambda q_iq_j}{\bq^4}, \nn \\ D^{i0}(q)&=& D^{0i}(q)=\frac{\lambda q^0 q^i}{\bq^4},\nn \\ D^{00} (q) &=& -\frac{1}{\bq^2}+\frac{\lambda q_0^2}{\bq^4}  \eea where $q=(q_0,\bq)$ and $q^2\equiv q_0^2-\bq^2$. These formulas simplify in the $\lambda\rw 0$ limit, wherein $D^{0i}=0$, thus $\ba$ and $a_0$ are decoupled, and the longitudinal modes of $\ba$ are eliminated\footnote{ The $\lambda\rw\infty$ limit is singular because $D^{\mu\nu}$ diverges.}.

What are
the meanings of the last three terms of Eq.(\ref{L-f-simp})? Let us first simplify the
$\frac{1}{2}(\bb')^2$ term. For notational simplicity, we define \bea
D(\bx,\by)\equiv \frac{1}{4\pi|\bx-\by|} \eea which satisfies \footnote{ In
this paper, $\nabla$ ( $\nabla'$ ) denotes differential operator with
respect to $\bx$ ($\by$ ).} $\bG(\bx,\by)=-\nabla D(\bx,\by)$.
With the input from Eq.(\ref{hprime}), we can readily obtain that (See Appendix \ref{sec:e-prime-square} for a derivation) \bea \frac{1}{2}\int d\bx
[\bb'(\bx)]^2 = \frac{g^2}{2}\int d\bx d\by \rho_m(\bx)
D(\bx,\by)\rho_m(\by) \label{b'square} \eea which is just the magnetic Coulomb
potential among monopoles. The $\frac{1}{2}(\be')^2 $ term in Eq.(\ref{L-f-simp}) can be simplified to current-current interactions (also see Appendix \ref{sec:e-prime-square}) \bea  \frac{1}{2}\int d\bx[\be'(\bx,t)]^2 =  \frac{g^2}{2}\int d\bx d\by D(\bx,\by) \bj_m^T(\bx,t) \cdot  \bj_m^T(\by,t) \label{e'square-simp} \eea where $\bj_m^T$ denotes the transverse part of $\bj_m$.

The last term of Eq.(\ref{L-f-simp}) is more interesting. It can be recast as (see Appendix \ref{sec:e-prime-square}) \bea  &&  -\int d\bx (\partial_t\ba )\cdot\be'  =
g \int d\bx d\by D(\bx,\by) [\nabla\times\partial_t\ba(\bx,t)] \cdot
\bj_m(\by,t)    \nn \\ &=&  g\int d\bx d\by
D(\bx,\by)\psi^\dag(\by,t)\balpha\cdot[\nabla\times\partial_t\ba(\bx,t)]\psi(\by,t)
\label{m-photon} \eea which describes a nonlocal
interaction between monopoles and electromagnetic field: $\ba(\bx,t)$ is
coupled to $\bj_m(\by,t)$ with a decaying factor $D(\bx,\by)$. It is apparently invariant under a gauge transformation of $\ba$ \footnote{A remark: In our formulation monopoles are inert to gauge transformation of $\ba$.}. In
momentum-frequency space, Eq.(\ref{m-photon})
gives rise to the monopole-photon interaction \bea g\frac{q_0}{\bq^2}
(\bq\times\ba_{q})\cdot\bj_m(-q)=
g\frac{q_0}{\bq^2}
\psi^\dag_{k+q}\balpha\cdot(\bq\times\ba_{q})
\psi_{k} \label{m-photon-2} \eea with $a_{q}$, $\bj_m(q)$, $\psi^\dag_{k}$ and
$\psi_{k}$ being the Fourier transformations of the
corresponding quantities in spacetime.
The ``$\bq\times\ba$'' factor in Eq.(\ref{m-photon-2}) indicates that
$\ba$ suffers a $\pi/2$ rotation around $\bq$ before being coupled to
monopoles.  This is physically intuitive, because $\bE$ and $\bB$ in
a propagating electromagnetic wave are related by a $\pi/2$ rotation
around $\bq$.

The monopole-photon interaction obtained in Eq.(\ref{m-photon-2}) is
apparently different from that of Ref.\cite{tu1978}, wherein the
interaction vertex does not contain a $q_0$ factor. In addition, the
direct current-current interaction, which is given by
Eq.(\ref{e'square-simp}), is absent in Ref.\cite{tu1978}. As far as
we can check for various physical processes, the two approaches lead
to the same results, though the present approach is more convenient
in many cases. In Appendix \ref{sec:equivalence}, we
return to a comparison of these two approaches.

Based on the above calculations, we present a more explicit expression for Eq.(\ref{L-f-simp}) as \bea && L_f +L_\lambda +L_g \nn \\
&=&\frac{1}{2} (-\partial_t\ba-\nabla a_0)^2 -
\frac{1}{2}(\nabla\times\ba)^2  +\frac{1}{2\lambda}(\nabla\cdot\ba)^2
\nn \\ && +   \psi^\dag(i\partial_t  + i\sum_{i=1}^3
\alpha_i\partial_i -\beta M)\psi \nn \\ && - \frac{g^2}{2}\int d\bx
d\by \rho_m(\bx,t) D(\bx,\by)\rho_m(\by,t) \nn \\ && +
\frac{g^2}{2}\int d\bx d\by D(\bx,\by) \bj_m^T(\bx,t) \cdot
\bj_m^T(\by,t)  \nn \\ && + g\int d\bx d\by D(\bx,\by)
[\nabla\times\partial_t\ba(\bx,t)]\cdot\bj_m(\by,t)
\label{monopole-explicit} \eea According to this Lagrangian, $a_0$ is
coupled only to the longitudinal part of $\ba$. Both $a_0$ and the
longitudinal part of $\ba$ are unphysical degrees of freedom. In
fact, $a_0$ has no dynamics of its own, thus, it can be
straightforwardly integrated out, resulting in a Lagrangian with only
the transverse part of $\ba$, which is denoted as $\ba^T$.
Integrating out $a_0$ is easiest in the $\lambda=0$ gauge. In this
gauge, the Maxwell equations tell us that $\nabla^2 a_0
=\nabla\cdot\bE =0$ (since electric charge is absent), therefore, we
have $a_0=$ constant, if the system has periodic boundary conditions.

Compared with the Hamiltonian approach\cite{tu1978}, in
which inspired guesswork ( about the monopole-photon interaction, etc ) is required, the present approach is more automatic.

Because of the unfamiliar form of monopole-photon interaction in the
above Lagrangian, some tests of its correctness are desirable. One of
the tests is as follows. We know that the Dirac equation for the electron
predicts that the electron has an intrinsic magnetic dipole moment with
$g$-factor $2$ (The ``$g$'' in ``$g$-factor'' should not be confused
with the magnetic coupling constant $g$ in our paper).  Since we take
magnetic monopoles as Dirac particles, we expect them to have an
intrinsic electric dipole moment, the ``electric $g$-factor'' taking
the same value ``$2$''. Now we are ready to calculate it in our
formalism\footnote{We would like to thank an anonymous referee for
suggesting us to calculate it.}. Following from the last term of
Eq.(\ref{monopole-explicit}), the matrix element between an initial
on-shell state $u(p)$ and a final on-shell state $u(p')$ of the
monopole is \bea g\frac{q_0}{|\bq|^2} u^\dag(p') [ (\bq\times
\ba_q)\cdot \balpha ] u(p)  \label{on-shell} \eea  Here it is useful
to recall the Gordon identity\cite{peskin1995,zee2010quantum} \bea
\bar{u}(p')\gamma^\mu u(p) =  \bar{u}(p')[\frac{(p'+p)^\mu}{2M}+
\frac{i\sigma^{\mu\nu}q_\nu}{2M}] u(p), \eea in which
$\gamma^0=\beta$, $\gamma^i = \alpha_i \beta$, $\bar{u}(p)
=u^\dag(p)\gamma^0$, $q_\mu=(p'-p)_\mu$, and
$\sigma^{\mu\nu}=\frac{i}{2}[\gamma^\mu, \gamma^\nu]$.  It is easy to see that, due to the second term at the right hand side of the Gordon
identity,  Eq.(\ref{on-shell}) gives rise to the following interaction
between monopole spin ${\bf s} =\frac{\bsigma}{2}$ and $\ba$ (in the nonrelativistic limit): \bea
\frac{ig}{2M}\frac{q_0}{|\bq|^2}(\bq\times\ba_q)\cdot(\bq\times\bsigma)
=  \frac{ig}{2M} q_0 \ba_q^T \cdot\bsigma =
\frac{g}{2M}\bE_q\cdot\bsigma, \eea where $\bE_q$ is the electric
field expressed in momentum space. This means that a fermionic
monopole described by the Dirac equation has an intrinsic electric
dipole moment $g\bsigma/2M$. The ``electric $g$-factor'' of the monopole
is $2$ \footnote{ Although our formalism is applicable to both
fermionic and bosonic monopoles, our calculation of electric moment
of monopole is specific to fermionic monopoles described by the Dirac
equation.}.

Monopole-photon systems can also be described by minimally coupling
monopoles to a (dual) gauge potential, the resultant theory being
equivalent to the usual quantum electrodynamics.  Therefore, the
formulation in this section can be regarded as another version of the
quantum electrodynamics\footnote{Electron-photon systems can also be
described in the formalism presented here, provided that we take
($c$) and ($d$) of the Maxwell equations as kinematic equations.}.
The merit of our formalism will manifest in its application to
electron-monopole-photon systems, to be investigated in the next
section.

\section{Interactions of electrons, magnetic monopoles, and photons}\label{sec:e-m-photon}

Having addressed the problem of monopole-photon interaction in the previous section, we shall formulate in this section the path integral quantization of electron-monopole-photon systems in the framework of fiber bundles, avoiding the troublesome ``string singularities''.

In addition to the monopole-photon interaction found in the previous section, for electron-monopole-photon systems we have to write down the electron-photon and electron-monopole interactions.  To this end, we express the kinematic part of magnetic
field as $\bb' =\nabla\times \ba'$, where $\ba'$ is given by \bea
\ba'(\bx,t) = g\int d\by\bA(\bx, \by)\rho_m(\by,t) \label{aprime} \eea
in which we have introduced a function $\bA(\bx,\by)$, which satisfies
$\nabla\times\bA(\bx,\by)=\bG(\bx,\by)$.
We can see that $\bA(\bx,\by)$ cannot be single-valued, otherwise, we
would have $\nabla\cdot(\nabla\times\ba') = \int d\by
\nabla\cdot[\nabla\times \bA(\bx, \by)] \rho_m(\by,t) = 0$. In fact,
the most natural language for this problem is the fiber
bundle\cite{wu1976,wu1976a}. Following Ref.\cite{wu1976,wu1976a}, we
divide the space into two overlapping patches, $R_a$ and $R_b$. It is
convenient to express $\bx-\by$ in the spherical coordinate
$(r,\theta,\phi)$ (with $\theta\in [0,\pi]$), namely, $\bx-\by=(r\sin\theta\cos\phi,r\sin\theta\sin\phi,r\cos\theta)$.  The first patch $R_a$
is defined by $\theta <\theta_0$, while the second patch $R_b$ is
defined by $\theta >\pi-\theta_0$,  $\theta_0$ being a constant in
$(\pi/2,\pi)$. We define\cite{wu1976a} \bea \bA_a(\bx,\by) &=&
\frac{1-\cos\theta}{4\pi r^2\sin^2\theta} \hat{\bz}\times(\bx-\by) \nn \\
\bA_b(\bx,\by) &=& -\frac{1+\cos\theta}{4\pi r^2\sin^2\theta}
\hat{\bz}\times(\bx-\by)   \label{Axy} \eea It is readily found that
$\bA_a(\bx,\by)-\bA_b(\bx,\by)=\frac{1}{2\pi}\nabla\phi$.

Let us define the full gauge potential \footnote{ Note
that ``$\bA(\bx,t)$'' and ``$\bA(\bx,\by)$'' refer to different
objects.} $\bA(\bx,t)=\ba(\bx,t)+\ba'(\bx,t)$ and $A_0(\bx,t)=a_0(\bx,t)+a'_0(\bx,t)$, where the mathematical formula for
$a'_0$ is to be determined shortly. The coupling of electromagnetic
field to electrons reads \bea && - e A_0(\bx,t) \rho_e(\bx,t)+ e\bA(\bx,t)\cdot\bj_e (\bx,t) \nn \\ &=& -e(a_0
+a'_0)\rho_e + e(\ba+\ba')\cdot\bj_e \eea  Including this
electromagnetic coupling,  we have the following Lagrangian density for Dirac
electrons \bea L_e = c^\dag (iD_t + i\sum_{i=1}^3\alpha_i D_i -\beta
m) c \eea where $c^\dag, c$ are the Grassmann numbers (anti-commuting
numbers) denoting electrons, $\alpha_i,\beta$ are the Dirac matrices,
$D_t=\partial_t + ie A_0$ and $D_i =\partial_i - ie A_i$ (or ${\bf
D}=\nabla - ie\bA$ ) are the covariant derivatives, and  $m$ is the
electron  (bare) mass.

Now we have to check the mathematical consistency of the definition
of $L_e$, and find the formula for $a'_0$.

Since $\ba$ is patch-independent, it is unambiguous, however, the
definition of $\ba'$ does depend on the patch choice.  In regions
where two patches overlap, it is unclear which patch we should choose
to define $\ba'$. Let us be more quantitative on this feature. Since
monopoles are point-like, the magnetic charge density can be written
as $g\rho_m(\by,t)=\sum_l g_l\delta(\by-\by_l)$  (We have $g_l=\pm
g$, where the minus sign is for anti-monopoles).   Each
(anti-)monopole labelled by $l$ determines two patches denoted as $a$
and $b$, in which $\bA(\bx,\by)$ in Eq.(\ref{aprime}) is defined as
$\bA_a(\bx,\by)$ and $\bA_b(\bx,\by)$, respectively. It is readily
seen that $\ba'$ in different patches are related by \bea
\ba'_{(a)}(\bx)-\ba'_{(b)}(\bx)  = \frac{1}{2\pi} g_l
\nabla\phi(\bx,\by_l) \label{patch-phase} \eea where $\by_l$ is the
position of monopole. To ensure the independence of $L_e$ on patch
choices,  the Grassmann numbers $c^\dag, c$ have to be
patch-dependent, and follow a prescribed transformation $c_{(b)}^\dag
=e^{-i\gamma} c_{(a)}^\dag, \, c_{(b)} = e^{i\gamma} c_{(a)}$, with
$\gamma$ suitably chosen. This is readily understood in the language
of fiber bundle. To keep $c^\dag (-i\sum_i\alpha_i D_i)c$ independent
on the patch choice, we find [using Eq.(\ref{patch-phase})] that
$\gamma$ satisfies $\nabla\gamma =-\frac{1}{2\pi} e  g_l
\nabla\phi(\bx,\by_l)$, which is equivalent to \bea
\exp(i\gamma)=\exp[-  i \frac{1}{2\pi}e g_l  \phi(\bx,\by_l)] \eea To
ensure that $e^{i\gamma}$ is single-valued, we must have the Dirac
quantization condition $eg/2\pi=$integer. In other words, the Dirac
quantization condition is necessary for $L_e$ to be independent of
the patch choice. Moreover, to preserve the patch-independence of
$c^\dag (iD_t)c$, we find that \bea a'^{(a)}_0 (\bx,t)-
a'^{(b)}_0(\bx,t) &=& -\frac{1}{2\pi}g_l\dot{\phi}(\bx,\by_l(t)) \nn \\
&=& \frac{1}{2\pi} g_l \dot{\by}_l(t)\cdot\nabla\phi(\bx,\by_l(t))
\eea To be consistent with this transformation rule, we have to
define \bea a'_0(\bx,t)= g\int d\by \bA(\bx,\by)\cdot\bj_m(\by,t)
\label{a'-zero} \eea With this definition we find that the equation
$\bE(\bx,t)=-\nabla A_0(\bx,t) -\partial_t\bA(\bx,t)$ is equivalent
to $\bE(\bx,t)= -\partial_t\ba(\bx,t) -\nabla
a_0(\bx,t)+\be'(\bx,t)$, which is a consistency check.

The full quantum theory is based on the path integral \bea \int D\ba
Da_0 Dc^\dag Dc D\psi^\dag D\psi\exp[i\int dt d^3x L(\bx,t)] \eea
where the Lagrangian density is \bea L &=& L_f+L_e+L_g + L_\lambda \nn \\ &=& \frac{1}{2}  (-\partial_t\ba-\nabla
a_0)^2 - \frac{1}{2}(\nabla\times\ba)^2
+\frac{1}{2\lambda}(\nabla\cdot\ba)^2   \nn \\  && +
\psi^\dag(i\partial_t + i\sum_{i=1}^3 \alpha_i\partial_i -\beta M)\psi
+c^\dag (iD_t +i\sum_{i=1}^3\alpha_i D_i -\beta m) c   \nn \\  &&
+\frac{1}{2} (\be')^2-\frac{1}{2}(\bb')^2  -(\partial_t\ba )\cdot\be'
\label{final} \eea Eq.(\ref{final}) is a central equation of the present paper. Now the dynamical part ( photon ) of electromagnetic field, described by $\ba$, is neatly separated out, while $\be'$ and $\bb'$ are
kinematic fields fully determined by electrons and monopoles[see
Eq.(\ref{hprime}) and Eq.(\ref{solution-eprime})]. Various
interactions are described by the covariant derivatives and the last
three terms of Eq.(\ref{final}). The monopole-photon interaction is automatically included in $L_f$, without the need to be put in by hand.

It is worth emphasizing that the
partial derivative $\partial_i$ appears in $L_g$, while the
covariant derivative $D_i$ appears in $L_e$. This feature resonates
with the classical Lagrangian theory\cite{wu1976}.

Taking advantage of Eq.(\ref{b'square}), Eq.(\ref{e'square-simp}), and Eq.(\ref{m-photon}), we have the following more explicit expression for Eq.(\ref{final})
\bea && \int d\bx L (\bx,t) = \frac{1}{2} \int d\bx ( -\partial_t\ba -\nabla a_0 )^2 - \frac{1}{2}\int d\bx (\nabla\times\ba )^2 \nn \\ && +  \frac{1}{2\lambda}\int d\bx (\nabla\cdot\ba)^2 + g \int d\bx d\by D(\bx,\by) [\nabla\times\partial_t\ba (\bx,t)] \cdot
\bj_m(\by,t)  \nn \\  && + \int d\bx
\psi^\dag(i\partial_t + i\sum_{i=1}^3 \alpha_i\partial_i -\beta M)\psi \nn \\ &&
+\int d\bx  c^\dag (iD_t + i\sum_{i=1}^3\alpha_i D_i -\beta m ) c   \nn \\  &&
-\frac{g^2}{2}\int d\bx d\by \rho_m(\bx,t)D(\bx-\by)\rho_m(\by,t) \nn \\ &&  + \frac{g^2}{2}\int d\bx d\by D(\bx,\by) \bj_m^T(\bx,t) \cdot  \bj_m^T(\by,t)
\label{final'} \eea
In the Coulomb gauge (the $\lambda = 0$ gauge), $a_0$ can be straightforwardly integrated out, yielding the electrical Coulomb energy. In this gauge the full Lagrangian can be recast as \bea && \int d\bx L (\bx,t) = \frac{1}{2} \int d\bx ( \partial_t\ba^T )^2 - \frac{1}{2}\int d\bx (\nabla\times\ba^T)^2 \nn \\ &&
+  g\int d\bx d\by D(\bx,\by) [\nabla\times\partial_t\ba^T(\bx,t)] \cdot
\bj_m^T(\by,t)  \nn \\  && + \int d\bx
\psi^\dag(i\partial_t + i\sum_{i=1}^3 \alpha_i\partial_i -\beta M)\psi \nn \\ &&
+\int d\bx \,  c^\dag [i(\partial_t + ie a'_0) + i\sum_{i=1}^3\alpha_i D_i -\beta m ] c   \nn \\  &&
-\frac{e^2}{2}\int d\bx d\by \,\rho_e(\bx,t)D(\bx-\by)\rho_e(\by,t) \nn \\ && -\frac{g^2}{2}\int d\bx d\by \, \rho_m(\bx,t)D(\bx-\by)\rho_m(\by,t) \nn \\ &&  + \frac{g^2}{2}\int d\bx d\by \, D(\bx,\by) \bj_m^T(\bx,t) \cdot  \bj_m^T(\by,t)
\label{final-coulomb} \eea where $\ba^T$ denotes the transverse part of $\ba$, and the longitudinal part of $\ba$ is absent. In the momentum space, the last three terms of Eq.(\ref{final-coulomb}) read \bea -\sum_\bq \frac{1}{2\bq^2} [e^2\rho_e(\bq,t)\rho_e(-\bq,t) + g^2 \rho_m(\bq,t)\rho_m(-\bq,t)]   \nn \\ + \sum_\bq \frac{g^2}{2\bq^2} ( \delta_{ij}  - \frac{q_i q_j }{ \bq^2 }) j^i_m(\bq,t)j^j_m(-\bq,t) \eea

\section{Equations of motion}\label{sec:motion}

We will first discuss equations of motion as extremal conditions in the variation method, then we proceed to promote them to quantum equations of motion, which is readily done in the path integral formulation.

We take the Lagrangian of electron-monopole-photon systems, namely Eq.(\ref{final'}), as our starting point. We now show that the Maxwell equations, and the Dirac equations and Lorentz equations for both electrons and monopoles can be obtained from variation method (We have to discard the gauge fixing term $L_\lambda$ in taking variation). Let us study the Maxwell equations first. The Maxwell equations $(a)$ and $(b)$ are automatically satisfied by our formulation, thus we only need to establish $(c)$ and $(d)$.

The action is defined as the integral of Lagrangian density, namely $S=\int dtd\bx\, L(\bx,t)$. The variational equation $ \delta S/\delta a_0  =0$ leads to \bea -\nabla\cdot(\partial_t\ba+\nabla a_0) -e\rho_e =0 \eea Because of $\nabla\cdot\be'=0$, this equation is equivalent to \bea    \nabla\cdot(-\partial_t\ba -\nabla a_0 +\be') = e\rho_e \eea which is simply the Maxwell equation $(c)$.

The variational equation $ \delta S/\delta \ba  =0$ leads to \bea -\partial_t (\partial_t \ba +\nabla a_0) -  \nabla\times (\nabla\times\ba ) +  \partial_t\be' + e\bj_e =0  \label{maxwell-d-1} \eea in which the first two terms are obvious, the $\partial_t\be'$ term comes from the fourth term of Eq.(\ref{final'}), and the $e\bj_e$ term comes from the sixth term of Eq.(\ref{final'}). Eq.(\ref{maxwell-d-1}) can be rewritten as  \bea  -\partial_t (-\partial_t \ba -\nabla a_0 +\be') +  \nabla\times ( \bb +\bb')   =   e\bj_e \label{maxwell-d-2} \eea  where we have used $\nabla\times\bb'=0$. Eq.(\ref{maxwell-d-2}) is simply the Maxwell equation $(d)$.

Now let us take the variation of the action with respect to $c^\dag$.
Because only the sixth term in Eq.(\ref{final'}) contributes, the
result is simply \bea (iD_t+i\sum_{i=1}^3 \alpha_i D_i -\beta m)c=0
\eea The remaining variation problem is $\delta S/\delta
\psi^\dag=0$. By a straightforward calculation, we have \bea   &&
\,\, (i\partial_t + i\balpha\cdot\bpartial-\beta M)\psi(\bx,t) \nn \\
&& - e g\int d\by \bA(\bx,\by)\cdot\balpha \,\psi(\bx,t)\rho_e(\by,t)
\nn \\&& +  e g\int d\by \bA(\bx,\by)\cdot\bj_e(\by,t)\,\psi(\bx,t)
\nn \\ &&  + g\int d\by D(\bx,\by)
[\nabla_\by\times\partial_t\ba(\by,t)]\cdot\balpha\,\psi(\bx,t) \nn
\\ && + g^2 \int d\by D(\bx,\by)\,\bj_m^T(\by,t)\cdot\balpha\,
\psi(\bx,t) \nn \\ &&   -g^2\int d\by
D(\bx,\by)\rho_m(\by,t)\,\psi(\bx,t) \nn \\ && = 0
\label{psi-variation} \eea which can be recast as \bea
(i\bar{D}_t+i\sum_{i=1}^3 \alpha_i \bar{D}_i -\beta M)\psi=0
\label{monopole-dirac} \eea where $\bar{D}_t=\partial_t +ig\bar{A}_0$
and $\bar{D}_i=\partial_i -ig\bar{A}_i$, with the potentials $\bar{A}_0$
and $\bar{\bA}$ being given by \bea \bar{A}_0(\bx,t) &=& -e\int d\by
\bA(\bx,\by)\cdot\bj_e(\by,t) +g\int d\by D(\bx,\by)\rho_m(\by,t),
\nn \\ \bar{\bA} (\bx,t) &=& -e\int d\by \bA(\bx,\by)\rho_e(\by,t) +
\int d\by D(\bx,\by) \nabla_\by\times\partial_t\ba(\by,t) \nn \\ && +
g\int d\by D(\bx,\by)\bj_m^T(\by,t)   \label{dual-A} \eea Although
the derivation of Eq.(\ref{monopole-dirac}) seems to be quite
straightforward in our formalism, it is useful to note a subtle point
here. Since $\bA(\bx,\by)$ depends on patch choice, the meaning of
Eq.(\ref{psi-variation}) looks ambiguous. To have a better
understanding, we notice that the part of the action [see
Eq.(\ref{final'})] that involves $\bA(\bx,\by)$ is symmetric between
electrons and monopoles, therefore, it is possible to construct the
patch structure according to the positions of electrons instead of
monopoles. With this slightly different interpretation, monopoles
undergo a gauge transformation $\psi_{(b)}=e^{i\gamma}\psi_{(a)}$ in
a patch switching, and the action given in Eq.(\ref{final'}) is still
independent on patch choice, thus, the action is unambiguous. It looks
safer to do $\delta S/\delta \psi^\dag$ with this interpretation,
because $c^\dag, c$ are patch-independent in this context. Now the
meaning of Eq.(\ref{psi-variation}) is quite understandable. We can
check that, if the Dirac quantization condition is satisfied, the
patch-dependence due to $\psi_{(b)}=e^{i\gamma}\psi_{(a)}$ cancels
the patch-dependence due to $\bA_a(\bx,\by)-\bA_b(\bx,\by)
=\frac{1}{2\pi}\nabla\phi$, thus, Eq.(\ref{psi-variation}) is
patch-independent.

On the other hand, the subtlety of equations of motion is not
fundamentally important here. The fundamental requirement is that the
action [Eq.(\ref{final'})], on which the path integral approach is
based, is unambiguous. This requirement is indeed satisfied. In our
formalism of electron-monopole-photon interaction, the meaning of the
action is more transparent than that of equations of motion.

In the path integral formalism, the classical equations of motion can be readily promoted to quantum equations of motion for the correlation functions of operators in the Heisenberg picture\cite{peskin1995}. Let us take the Dirac equation for electrons as an example. By the invariance of the path integral \bea  \int
D\ba Da_0 Dc^\dag Dc D\psi^\dag D\psi \, \exp(iS)\, c^\dag (\bx_1,t_1) \eea under a shifting of electron variable $c^\dag(\bx,t)$, we can obtain that \bea  \int
D\ba Da_0 Dc^\dag Dc D\psi^\dag D\psi \, \exp(iS) \nn \\ \times [i\frac{\delta S}{\delta c(\bx,t)} c^\dag (\bx_1,t_1) +\delta(\bx-\bx_1)\delta(t-t_1)] =0\eea
According to the general correspondence between the path integral (with insertions of field variables at arbitrary spacetime points) and the correlation functions of operators in the Heisenberg picture (see Ref.\cite{peskin1995}), we have
\bea  \la\Omega| T(iD_t + i\sum_{i=1}^3\alpha_i D_i -\beta m)\,c_H(\bx,t)  c_H^\dag(\bx_1,t_1) |\Omega\ra \nn \\ = -i\delta(\bx-\bx_1)\delta(t-t_1) \label{electron-quantum}  \eea where $T$ denotes time ordering, $|\Omega\ra$ denotes the vacuum state (or ground state), and the subscript ``$H$'' refers to the Heisenberg picture. The simple equation $\delta S/\delta c^\dag(\bx,t)= (iD_t + i\sum_{i=1}^3\alpha_i D_i -\beta m)\,c(\bx,t)$ has been used in deriving Eq.(\ref{electron-quantum}). More concisely, we can write down the operator equation \bea (iD_t + i\sum_{i=1}^3\alpha_i D_i -\beta m)\,c_H(\bx,t) =0\eea In this way classical equations of motion can be translated into operator equations. Similarly, the Dirac equation for monopoles, and the Maxwell equations can be translated into operator equations.

Now let us study the fate of Lorentz equations. In classical electrodynamics, an electron feels Lorentz force in an electromagnetic field, such that its momentum satisfies $\dot{\bp} =e(\bE+\bv\times\bB)$, where $\bv$ is the velocity of the electron. Similarly, a monopole satisfies a dual Lorentz equation $\dot{\bp} = g(\bB-\bv\times \bE)$, in which $\bv$ is the velocity of monopole. We would like to find the counterparts of classical Lorentz equations in our formulation.

First we study the Lorentz equation for electrons. The momentum $ p_i$ in the classical Lorentz equation is replaced by the local operator $c_H^\dag (-i D_i) c_H(\bx,t)$, and we have the operator equation \bea && \frac{d}{dt} [ c_H^\dag (-i D_i) c_H ] \nn \\ &=& \frac{dc_H^\dag}{dt} (-i D_i) c_H + c_H^\dag (-i D_i)\frac{dc_H}{dt}  + c_H^\dag (-i \frac{d  D_i}{dt} ) c_H \nn \\ &=& -c^\dag_H [D_i, eA_0- i\balpha\cdot\bD +\beta m ] c_H  -ec^\dag_H (\partial_t A_i) c_H \nn \\ &=&   e \epsilon_{ijk} (c^\dag_H\alpha_j c_H) B_k - e(\partial_i A_0) c^\dag_H c_H - e (\partial_t  A_i) c^\dag_H c_H \nn \\ &=&   e \epsilon_{ijk}  j_e^j B_k - e E_i \rho_e \eea or equivalently, \bea \frac{d}{dt} [ c_H^\dag (-i \bD) c_H ] = e\rho_e \bE + e\bj_e\times\bB \eea where $\bE$ and $\bB$ are also understood as operators in the Heisenberg picture. This is the Lorentz equation in our formulation. It is a local operator equation. Similarly, we can obtain a Lorentz equation for monopoles, which reads \bea \frac{d}{dt} [ \psi_H^\dag (-i \bD) \psi_H ] = g\rho_m \bar{\bE} + g\bj_m\times\bar{\bB} \label{lorentz-monopole-1} \eea where $\bar{\bE} = -\partial_t\bar{\bA} -\nabla\bar{A}_0$ and $\bar{\bB} =\nabla\times\bar{\bA}$. In the Appendix \ref{sec:dual-calculation} we show that \bea \bar{\bB}(\bx,t)  = -\bE(\bx,t),  \label{dual-B} \eea
and
\bea \bar{\bE}(\bx,t)= \bB(\bx,t)  \label{dual-E} \eea therefore, Eq.(\ref{lorentz-monopole-1}) can be recast as \bea \frac{d}{dt} [ \psi_H^\dag (-i \bD) \psi_H ] = g\rho_m  \bB - g\bj_m\times\bE \label{lorentz-monopole} \eea To summarize this section, we have established that all Maxwell equations and Lorentz equations hold in our formulation as operator equations. The efficiency of promoting classical equations of motion to operator equations in the path integral formalism is notable.

\section{Dual formulation}\label{sec:dual}

In Sec.\ref{sec:motion} we implicitly touched the dual description, where the dual electromagnetic fields $\bar{\bB}$ and $\bar{\bE}$ were used.  To discuss the dual formulation in a transparent way, we define the dual quantities \bea \bar{e} = g, \,\, \bar{g}= -e, \\ \bar{\rho}_e = \bar{e}\psi^\dag\psi = \rho_m, \, \bar{\rho}_m =  c^\dag c=  \rho_e, \\ \bar{\bj}_e = \psi^\dag\balpha\psi = \bj_m, \, \bar{\bj}_m =  c^\dag\balpha c = \bj_e,  \\ \bar{\bB}= -\bE, \, \bar{\bE} = \bB, \label{dual-B-E} \eea
and the covariant derivatives \bea  \bar{D}_t &=& \partial_t +i\bar{e}\bar{A}_0 = \partial_t +i\bar{e}(\bar{a}'_0+\bar{a}_0), \nn \\  \bar{D}_i &=& \partial_i -i\bar{e}\bar{A}_i  =\partial_i -i\bar{e}(\bar{a}'_i+ \bar{a}_i )  \eea  The dual equations of Eq.(\ref{EH}) read \bea \bar{\bE} = -\partial_t\bar{\ba} -\nabla\bar{a}_0 +\bar{\be}'; \, \bar{\bB} = \nabla\times\bar{\ba}+\bar{\bb}' \label{dual-of-EH} \eea in which  $\bar{\be}'$ is defined as  \bea \bar{\be}'(\bx,t)= \bar{g}\int d\by
\bG(\bx,\by)\times \bar{\bj}_m(\by,t) \nn \\ = -e \int d\by
\bG(\bx,\by)\times \bj_e(\by,t) \eea which is dual to  Eq.(\ref{solution-eprime}). Similarly, $\bar{\bb}'$ is defined as the dual equation of Eq.(\ref{hprime}). Together with the relation $\bar{\bB}=-\bE$, Eq.(\ref{dual-of-EH}) implies  \bea \bar{\ba} (\bx,t) =  \int
d\by D(\bx,\by) \nabla_\by\times\partial_t\ba(\by,t) + g\int d\by
D(\bx,\by)\bj_m^T(\by,t),  \label{dual-a} \eea which is also suggested by
Eq.(\ref{dual-A}). Similarly, we also have\footnote{For simplicity we take the Coulomb gauge ($\lambda=0$) in this section, therefore, $\ba$ contains only transverse modes.}  \bea \ba (\bx,t) =  -\int
d\by D(\bx,\by) \nabla_\by\times\partial_t \bar{\ba}(\by,t) + e\int d\by
D(\bx,\by)\bj_e^T(\by,t)  \label{dual-a-bar} \eea As a consistency check, we can see that Eq.(\ref{dual-a-bar}) can be obtained from Eq.(\ref{dual-a}) by adding an overbar to each variable and using $\bar{\bar{\ba}}=-\ba$.

In the remainder of this section, we would like to show that a dual description can be obtained by changing the variables of the path integral from $(a_0, \ba)$ to $(\bar{a}_0, \bar{\ba})$. Furthermore, we show that the dual description is equivalent to the original description.
For the purpose of this section, it is convenient to use a more compact but equivalent expression for the Lagrangian density, which reads  \bea L &=& \frac{1}{2}(\bE^2 -\bB^2)  \nn \\  &&
+\psi^\dag (i\partial_t+i\balpha\cdot\bpartial-\beta M)\psi + c^\dag (i\partial_t+i\balpha\cdot\bpartial-\beta m)c \nn \\ && -e(a_0+a'_0)\rho_e +e(\ba +\ba')\cdot\bj_e + L_\lambda, \label{L-1} \eea in which $\bE$ and $\bB$ are given by Eq.(\ref{EH}). It is readily seen that the Lagrangian density given in Eq.(\ref{L-1}) is equal to the one given in Eq.(\ref{final}). For simplicity we take the $\lambda\rw 0$ limit (the Coulomb gauge) in this section, such that only transverse modes of $\ba$ remain.

In the dual description, we use the dual gauge potentials $\bar{\ba}$ and $\bar{a}_0$ as the fundamental variables in the path integral.  The dual Lagrangian $\bar{L}$ is obtained from $L$ by simply adding an overbar to each electromagnetic quantity. It is given as \bea \bar{L} &=& \frac{1}{2}(\bar{\bE}^2 -\bar{\bB}^2)  \nn \\  &&
+\psi^\dag (i\partial_t+i\balpha\cdot\bpartial-\beta M)\psi + c^\dag (i\partial_t+i\balpha\cdot\bpartial-\beta m)c \nn \\ && -\bar{e}(\bar{a}_0+\bar{a}'_0)\bar{\rho}_e + \bar{e}(\bar{\ba} +\bar{\ba}')\cdot\bar{\bj}_e +\bar{L}_\lambda, \label{L-2} \eea  in which the gauge fixing term in Eq.(\ref{L-2}) is given as  $\bar{L}_\lambda=\frac{1}{2\lambda}(\nabla\cdot\bar{\ba})^2$.

It is a straightforward exercise to expand the dual Lagrangian $\bar{L}$ as \bea \bar{L} &=& \frac{1}{2}  (-\partial_t\bar{\ba}-\nabla
\bar{a}_0)^2 - \frac{1}{2}(\nabla\times\bar{\ba})^2
+\frac{1}{2\lambda}(\nabla\cdot\bar{\ba})^2   \nn \\  && +
\psi^\dag (i\bar{D}_t +i\sum_{i=1}^3\alpha_i \bar{D}_i -\beta M) \psi
+c^\dag (i\partial_t + i\sum_{i=1}^3 \alpha_i\partial_i -\beta m) c   \nn \\  &&
+\frac{1}{2} (\bar{\be}')^2-\frac{1}{2}(\bar{\bb}')^2  -(\partial_t\bar{\ba} )\cdot\bar{\be}'
\label{final-dual} \eea Compared to Eq.(\ref{final}), the covariant derivative in Eq.(\ref{final-dual}) is associated with monopoles instead of electrons.

Now we would like to show that the difference $L - \bar{L}$, without
inclusion of the gauge fixing terms $L_\lambda$ and
$\bar{L}_\lambda$, is actually a total derivative, therefore, the two
Lagrangians $\int d\bx L(\bx,t)$ and $\int d\bx \bar{L}(\bx,t)$ are
equivalent. In fact, with the input of Eq.(\ref{dual-B-E}), we have \bea && \int d\bx  L- \int d\bx \bar{L} =
\int d\bx [\bE^2 -\bB^2 -e a_0\rho_e +e \ba\cdot\bj_e \nn \\ && +
\bar{e}\bar{a}_0\bar{\rho}_e - \bar{e}\bar{\ba}\cdot\bar{\bj}_e +
(e\ba'\cdot\bj_e -ea'_0\rho_e -\bar{e}\bar{\ba}'\cdot\bar{\bj}_e +
\bar{e}\bar{a}'_0\bar{\rho}_e )]  \eea in which we have excluded the
gauge fixing terms $L_\lambda$ and $\bar{L}_\lambda$. It is not
difficult to check that the last four terms in the parenthesis
vanish. Moreover, the Maxwell equations can be used to rewrite $e\int d\bx
(-a_0\rho_e +\ba\cdot\bj_e)$ as \bea  && e\int d\bx  (-a_0\rho_e
+\ba\cdot\bj_e)   \nn \\ &=&  \int d\bx  [-a_0\nabla\cdot\bE +
(\nabla\times\bB-\partial_t\bE)\cdot\ba ]    \nn \\ &=&  \int d\bx
(-\bE^2 +\bB^2 +\bE\cdot\be' -\bB\cdot\bb') +\partial_t(\dots) \eea
in which ``$\partial_t(\dots)$'' denotes total derivatives with
respect to $t$. It follows that \bea  && \int d\bx  L- \int d\bx
\bar{L} \nn \\ &=& \int d\bx ( \bE\cdot\be' -\bB\cdot\bb' +
\bar{e}\bar{a}_0\bar{\rho}_e - \bar{e}\bar{\ba}\cdot\bar{\bj}_e )
+\partial_t(\dots)    \eea In the Coulomb
gauge ($\lambda=0$) in use, we have $\bar{e}\int d\bx\,
\bar{a}_0\bar{\rho}_e = \int d\bx\,\bar{a}_0 \nabla\cdot\bB = -\int
d\bx\,\nabla\bar{a}_0 \cdot\bB= \int d\bx\, \bb' \cdot\bB$,
therefore, we have \bea  && \int d\bx  L- \int d\bx \bar{L} \nn \\
&=& \int d\bx ( \bE\cdot\be'   -\bar{e}\bar{\ba}\cdot\bar{\bj}_e )
+\partial_t(\dots) \label{dual-simp} \eea in which the first term can
be recast, according to Eq.(\ref{EH}), as \bea && \int d\bx  \,
\bE\cdot\be' = -g\int d\bx d\by \, D(\bx,\by)
\bj_m(\bx)\cdot[\nabla_\by \times\bE(\by)]  \nn \\ &=&  g^2\int d\bx
d\by d\bz \, D(\bx,\by) \bj_m(\bx)\cdot\{ \nabla_\by\times
[\nabla_\by\times  (D(\by,\bz)\bj_m(\bz,t)) ] \} \nn \\ && + g\int
d\bx d\by \, D(\bx,\by) \bj_m(\bx)\cdot [\nabla_\by\times \partial_t
\ba(\by,t)], \label{sum-1}  \eea In addition, we can make use of Eq.(\ref{dual-a}) and rewrite the second term in
Eq.(\ref{dual-simp}) as \bea  && -\bar{e}\int
d\bx\,\bar{\ba}\cdot\bar{\bj}_e  = -g\int d\bx d\by D(\bx,\by)
\bj_m(\bx) \cdot [ \nabla_\by\times\partial_t\ba(\by,t)]   \nn \\ &&
- g^2\int d\bx d\by D(\bx,\by)\bj_m^T(\bx)\cdot\bj_m^T(\by,t)
\label{sum-2} \eea By summing Eq.(\ref{sum-1}) and Eq.(\ref{sum-2}),
it is now straightforward to see that Eq.(\ref{dual-simp}) reads \bea
\int d\bx  L- \int d\bx \bar{L}  = \partial_t(\dots) \eea which is
the central result of this section.

In the original description with Lagrangian $L$ given in Eq.(\ref{final}) or Eq.(\ref{L-1}), electron-photon interaction is apparent in the covariant derivative, while the monopole-photon interaction comes from $L_f=\frac{1}{2}(\bE^2-\bB^2)$. In the dual description with Lagrangian given in Eq.(\ref{L-2}) or Eq.(\ref{final-dual}), monopole-photon interaction is apparent in the covariant derivative, while electron-photon interaction comes from $\bar{L}_f=\frac{1}{2}(\bar{\bE}^2-\bar{\bB}^2)$. It is assuring to see in this section that $L=\bar{L} \, +$ total derivative terms, therefore, $L$ and $\bar{L}$ describe the same physics, as they should do. The interested readers can also read Ref.\cite{wu1976} for the dual transformation of the classical Lagrangian.

\section{Effective monopole-monopole interaction: A consistency check of the proposed Lagrangian}\label{sec:consistent}

In our formulation, monopoles are coupled to electromagnetic field  in a unusual manner. For instance, there is a $q_0$ factor in the monopole-photon interaction found in Eq.(\ref{m-photon-2}). In Sec.\ref{sec:m-photon} we have calculated the electric moment of a Dirac monopole, and the result is exactly what we expect. This is a nontrivial test of our formalism. In this section, we would like to design more tests for our formalism.

For simplicity of notations, let us consider monopole-photon systems without the complication of electrons. We would like to calculate the effective action of monopoles after photons are integrated out. There are two methods to do this, as given in Sec.\ref{sec:approach-1} and Sec.\ref{sec:approach-2} below, which, by the internal consistency of our formulation, should lead to the same result.

\subsection{Effective monopole-monopole interaction in the dual description}\label{sec:approach-1}

This method is the easier one.  Because of electromagnetic duality, the monopole-photon problem is equivalent to the electron-photon problem. In
other words, we can regard the monopole-photon problem as the dual of
quantum electrodynamics. In this approach, magnetic charges are
minimally coupled to the electromagnetic field as \bea
\bar{e}(\bar{\ba}\cdot\bar{\bj}_e - \bar{a}_0 \bar{\rho}_e), \eea in
which $\bar{\bj}_e \equiv\bj_m$, $\bar{\rho}_e= \rho_m$, and
$\bar{e}=g$. The quantities with overbar are the dual variables (see
Sec.\ref{sec:dual}). It is straightforward to integrate out photons
in the Coulomb gauge, yielding the effective action for monopoles as
\bea S_{{\rm eff}} &=& \int dt d\bx L_g - \frac{\bar{e}^2}{2} \sum_q
\frac{1}{q^2} (\delta_{ij} - \frac{q_i q_j}{\bq^2} )
\bar{j}_e^i(q)\bar{j}_e^j(-q)  \nn \\ && - \frac{\bar{e}^2}{2}\sum_q
\frac{1}{\bq^2}\bar{\rho}_e (q)\bar{\rho}_e (-q),    \eea or
equivalently, \bea S_{{\rm eff}} &=& \int dt d\bx L_g -\frac{g^2}{2}
\sum_q \frac{1}{q^2} (\delta_{ij} - \frac{q_i q_j}{\bq^2} )
j_m^i(q)j_m^j(-q)  \nn \\ && - \frac{g^2}{2}\sum_q
\frac{1}{\bq^2}\rho_m (q)\rho_m (-q)   \label{method-1} \eea where
$q=(q_0,\bq)$, and $q^2\equiv q_0^2-\bq^2$. The second term describes
current-current interaction, and the last term describes the magnetic
Coulomb energy.

\subsection{Effective monopole-monopole interaction in the original description}\label{sec:approach-2}

In the original description, monopole-monopole interaction is described by the Lagrangian density given in Eq.(\ref{final'}).  There are two contributions to the effective magnetic current-current interaction. The first part is mediated by $\ba$, the interaction vertex being given by Eq.(\ref{m-photon-2}). Its contribution to the effective action of monopoles is found to be \bea   - \frac{g^2}{2} \sum_q \frac{1}{q^2} \frac{q_0^2}{\bq^2} (\delta_{ij} - \frac{q_i q_j}{\bq^2} ) j_m^i(q)j_m^j(-q),  \eea which looks quite different from the current-current interaction in Eq.(\ref{method-1}), because of the awkward $ q_0^2/\bq^2$ factor. Fortunately, there is a second contribution to the current-current interaction, namely the $\frac{1}{2}(\be')^2$ term, which is simplified in Eq.(\ref{e'interaction}). Adding these contributions together, we have the total current-current interaction \bea  &&  - \frac{g^2}{2} \sum_q \frac{1}{q^2} \frac{q_0^2}{\bq^2} (\delta_{ij} - \frac{q_i q_j}{\bq^2} ) j_m^i(q)j_m^j(-q)   \nn \\ && +  \frac{g^2}{2} \sum_{q} \frac{1}{\bq^2} ( \delta_{ij} - \frac{q_i q_j }{ \bq^2 }) j^i_m(q)j^j_m(-q)  \nn \\ &=& - \frac{g^2}{2} \sum_q (\frac{1}{q^2} \frac{q_0^2}{\bq^2} -\frac{1}{\bq^2}) (\delta_{ij} - \frac{q_i q_j}{\bq^2} ) j_m^i(q)j_m^j(-q) \nn \\ &=& - \frac{g^2}{2} \sum_q \frac{1}{\bq^2}  \frac{q_0^2-q^2}{ q^2}   (\delta_{ij} - \frac{q_i q_j}{\bq^2} ) j_m^i(q)j_m^j(-q) \nn \\ &=& - \frac{g^2}{2} \sum_q \frac{1}{q^2}   (\delta_{ij} - \frac{q_i q_j}{\bq^2} ) j_m^i(q)j_m^j(-q)  \eea This current-current interaction is the same as the one found in Eq.(\ref{method-1}). The awkward $q_0$ factor turns out to be an indispensable part of the entire theory. The Coulomb energy is given by the $-\frac{1}{2}(\bb')^2$ term in the Lagrangian density, and also equals to the last term of Eq.(\ref{method-1}).  Therefore, the complete effective action takes the same form as Eq.(\ref{method-1}).

This exact match between two vastly different approaches reinforces our confidence in the validity of Eq.(\ref{final}).

As a final remark to this section, we mention that the effective electron-monopole interaction can also be found in the original and dual descriptions, with matching results. Without going into details, we note that the effective electron-monopole interaction mediated by photon takes the form of $eg q_0\bq\cdot[\bj_e(q)\times\bj_m(-q)]/|\bq|^2( q_0^2-|\bq|^2)$. This part of the effective electron-monopole interaction can also be obtained using the Hamiltonian formalism\cite{tu1978}, though the present approach is more convenient (e.g. The appearance of $q_0$ factor is less straightforward in the approach of Ref.\cite{tu1978}).

\section{Final remarks}

In this paper we have formulated a method for the quantization of
electron-monopole-photon systems through the path integral approach.
In this formulation, the electron-photon, monopole-photon, and
electron-monopole interactions emerge in a natural manner, for
instance, the monopole-photon interaction is automatically generated
from the Lagrangian of electromagnetic fields. In our formulation no
Dirac string is involved, thanks to the language of fiber bundle.

Our formulation is applicable in both relativistic and
nonrelativistic cases. Since the Coulomb gauge is used, Lorentz
invariance is not manifest in this formulation.

On the one hand, the interaction of electrons, magnetic monopoles, and photons is a fundamental theoretical topic. On the other hand,
magnetic monopoles have long been candidates of fundamental particles in high energy physics.
Recently monopoles have also found renewed interest
in condensed matter physics\cite{ladak2010,jaubert2009,qi2009,morris2009, castelnovo2008,kadowaki2009,bramwell2009}. We thus hope that our approach will be useful to the study of a variety of systems.

In the present paper we have not addressed the problem of
renormalization, which is left for future works. Finally, we would
like to remark that calculations involving nontrivial fiber bundles
are necessarily subtle, though our formalism provides a transparent
starting point. The subtlety should be properly handled to obtain
correct results.

\section{Acknowledgements}

I am deeply grateful to Professor Chen Ning Yang for bringing my
attention to this problem, for many illuminating discussions, and for
his encouragement during this work. The author is supported by NSFC
under Grant No. 11304175 and Tsinghua University Initiative
Scientific Research Program.

\appendix

\section{Derivation of Eq.(\ref{solution-eprime})}\label{sec:derivation1}

In this appendix we show that Eq.(\ref{solution-eprime}) is a
solution of Eq.(\ref{eprime}). Taking the curl of
Eq.(\ref{solution-eprime}), we have \bea  && \nabla\times \be'(\bx,t)
\nn \\ &=& g\nabla\times [\int d\by \bG(\bx,\by)\times \bj_m(\by,t)]
\nn \\ &=& -g\int d\by\bj_m(\by,t)\nabla\cdot\bG(\bx,\by) + g\int
d\by [\bj_m(\by,t)\cdot\nabla]\bG(\bx,\by) \nn \\ &=& -g\int
d\by\bj_m(\by,t)\delta(\bx-\by) -  g\int d\by
[\bj_m(\by,t)\cdot\nabla']\bG(\bx,\by) \nn \\ &=& - g\bj_m(\bx,t) +
g\int d\by  [\nabla'\cdot\bj_m(\by,t)] \bG(\bx,\by) \nn \\ &=&
-g\bj_m(\bx,t) - g\int d\by \partial_t  \rho_m(\by,t)\bG(\bx,\by) \nn
\\ &=& -g\bj_m(\bx,t) - g\partial_t [ \int d\by
\rho_m(\by,t)\bG(\bx,\by)] \label{curl-e'} \eea where ``$\nabla$ (
$\nabla'$ )'' refers to the gradient with respect to $\bx$ ( $\by$ ).
In this calculation we have used the law of conservation of  magnetic
charge. Combining Eq.(\ref{curl-e'}) with the definition of $\bb'$,
namely Eq.(\ref{hprime}), we have \bea \nabla\times \be'(\bx,t) +
\partial_t \bb'(\bx,t) = -g\bj_m(\bx,t) \eea which is just
Eq.(\ref{eprime}). We also note that $\nabla\cdot\be'=0$.

\section{Simplification of Eq.(\ref{L-f-simp}) }\label{sec:e-prime-square}

We have recast the last three terms of Eq.(\ref{L-f-simp}) as the expressions presented in Eq.(\ref{b'square}), Eq.(\ref{e'square-simp}), and Eq.(\ref{m-photon}). In this appendix we provide calculational details.

First let us provide a derivation for Eq.(\ref{b'square}), which is not difficult. Using Eq.(\ref{hprime}), we have
\bea \frac{1}{2}\int d\bx
[\bb'(\bx)]^2 &=& \frac{g}{2}\int d\bx d\by \bb'(\bx)\cdot
\bG(\bx,\by)\rho_m(\by) \nn \\ &=& -  \frac{g}{2}\int d\bx d\by
\bb'(\bx)\cdot  \nabla D(\bx,\by)\rho_m(\by)   \nn \\ &=&
\frac{g}{2} \int d\bx d\by \nabla\cdot \bb'(\bx) D(\bx,\by)\rho_m(\by)
\nn \\ &=& \frac{g^2}{2}\int d\bx d\by \rho_m(\bx)
D(\bx,\by)\rho_m(\by) \eea which is just Eq.(\ref{b'square}).

Next we shall derive Eq.(\ref{e'square-simp}). By a somewhat tedious calculation\footnote{
In the above calculation we have used well known formulas in vector analysis, such as $\nabla\cdot({\bf X}\times{\bf Y})={\bf Y}\cdot(\nabla\times{\bf X}) - {\bf X}\cdot(\nabla\times{\bf Y})$, where ${\bf X},{\bf Y}$ are continuous vector functions.}, we have \bea && \frac{1}{2}\int d\bx[\be'(\bx,t)]^2 \nn \\ &=& \frac{g^2}{2}\int d\bx d\by d\bz [\bG(\bx,\by)\times \bj_m(\by,t)]\cdot[\bG(\bx,\bz)\times \bj_m(\bz,t)] \nn \\ &=& \frac{g^2}{2}\int d\bx d\by d\bz [\nabla D(\bx,\by)\times \bj_m(\by,t)]\cdot[\nabla D(\bx,\bz)\times \bj_m(\bz,t)] \nn \\ &=& \frac{g^2}{2}\int d\bx d\by d\bz \{ \nabla\times [ D(\bx,\by) \bj_m(\by,t)]\} \cdot \{ \nabla\times [ D(\bx,\bz) \bj_m(\bz,t)]\} \nn \\
&=& \frac{g^2}{2}\int d\bx d\by d\bz   D(\bx,\bz) \bj_m(\bz,t) \cdot \{ \nabla\times [ \nabla\times ( D(\bx,\by) \bj_m(\by,t) )]\}  \nn \\  &=& \frac{g^2}{2}\int d\bx d\by d\bz   D(\bx,\bz) \bj_m(\bz,t) \cdot \{ \nabla [ \nabla\cdot  ( D(\bx,\by) \bj_m(\by,t) )]  \nn \\ && -\nabla^2 [D(\bx,\by)\bj_m(\by,t)] \}  \nn \\  &=& \frac{g^2}{2}\int d\bx d\by d\bz   D(\bx,\bz) \bj_m(\bz,t) \cdot \{ \nabla [ \nabla  D(\bx,\by) \cdot\bj_m(\by,t) )]  \nn \\ && + \delta(\bx-\by)\bj_m(\by,t)  \}
\nn \\  &=&  \frac{g^2}{2}\int d\bx d\by D(\bx,\by) \bj_m(\bx,t) \cdot  \bj_m(\by,t) \nn \\ && +  \frac{g^2}{2}\int d\bx d\by d\bz   D(\bx,\bz) \bj_m(\bz,t) \cdot \{ \nabla [ \nabla  D(\bx,\by) \cdot\bj_m(\by,t) )]   \}
\nn \\  &=&  \frac{g^2}{2}\int d\bx d\by D(\bx,\by) \bj_m(\bx,t) \cdot  \bj_m(\by,t) \nn \\ && -  \frac{g^2}{2}\int d\bx d\by d\bz   \{ \nabla \cdot[ D(\bx,\bz) \bj_m(\bz,t) ]\}  [ \nabla  D(\bx,\by) \cdot\bj_m(\by,t) )]
\nn \\  &=&  \frac{g^2}{2}\int d\bx d\by D(\bx,\by) \bj_m(\bx,t) \cdot  \bj_m(\by,t) \nn \\ && -  \frac{g^2}{2}\int d\bx d\by d\bz   [ \nabla  D(\bx,\by) \cdot\bj_m(\by,t) )][\nabla  D(\bx,\bz) \cdot \bj_m(\bz,t) ] \nn \\ &=& \frac{g^2}{2}\int d\bx d\by D(\bx,\by)   \bj_m^T(\bx,t) \cdot \bj_m^T (\by,t)
\label{current-interaction} \eea where $\nabla$ denotes differential operators with respect to the coordinate $\bx$. The expression $\bj_m^T(\bq) $ denotes the transverse part of $\bj_m(\bq)$. In momentum space, $\bj^T_m$ is defined by \bea \bj_m^T(\bq) \equiv \frac{[\bq\times\bj_m(\bq)]\times\bq}{|\bq|^2} = \bj_m(\bq) -\frac{ \bq\cdot\bj_m(\bq)}{|\bq|^2}\bq \eea or \bea j_m^{T,i}(\bq) \equiv (\delta_{ij}-\frac{q_i q_j}{|\bq|^2})j_m^j(\bq) \eea where $j_m^i$ denotes the $i$-th component of $\bj_m$, and similarly for $j^{T,i}_m$.

Eq.(\ref{current-interaction}) can be recast in momentum space as  \bea  && \frac{1}{2}  \int d\bx[\be'(\bx,t)]^2 \nn \\ &=& \frac{g^2}{2} \sum_{\bq}   j^{T,i}_m(\bq,t)D_{ij}(\bq) j^{T,j}_m(-\bq,t)     \nn \\ &=& \frac{g^2}{2} \sum_{q} \frac{1}{\bq^2} ( \delta_{ij} - \frac{q_i q_j }{ \bq^2 }) j^{i}_m(\bq)j^{j}_m(-\bq)  \label{e'interaction} \eea

We would like to mention that the last line of Eq.(\ref{current-interaction}) can also be written as $\frac{1}{2}\int d\bx d\by  j_m^i(\bq)D_{ij}^T(\bq) j_m^j(\bq)$, where we have defined the shorthand notation \bea D^T_{ij}(\bx,\by)  = D(\bx,\by)\delta_{ij} -\int d\bz \frac{\partial}{\partial z_i} D(\bz,\bx)\frac{\partial}{\partial z_j} D(\bz,\by) \eea  In momentum space we have \bea D^T_{ij}(\bq) = \frac{1}{\bq^2} ( \delta_{ij} - \frac{q_i q_j }{ \bq^2 }) \eea

The last term in Eq.(\ref{L-f-simp}) has been simplified to the expression given in Eq.(\ref{m-photon}).  The detail of this calculation is given as follows
\bea &&  -\int d\bx (\partial_t\ba )\cdot\be' =   -\int d\bx  d\by
[\partial_t\ba(\bx,t)]\cdot  [\bG(\bx,\by)\times \bj_m(\by,t)] \nn \\ &=&  \int d\bx d\by
[\partial_t\ba(\bx,t)]\cdot \{\nabla\times [D(\bx,\by)\bj_m(\by,t)]\} \nn
\\ &=&  -\int d\bx d\by  \nabla\cdot
[\partial_t\ba(\bx,t)\times\bj_m(\by,t)D(\bx,\by)] \nn \\ && + \int d\bx d\by
D(\bx,\by)\bj_m(\by,t)\cdot[\nabla\times\partial_t\ba(\bx,t)] \nn \\ &=&
 \int d\bx d\by D(\bx,\by) [\nabla\times\partial_t\ba(\bx,t)] \cdot
\bj_m(\by,t)    \nn \\ &=&  g\int d\bx d\by
D(\bx,\by)\psi^\dag(\by,t)\balpha\cdot[\nabla\times\partial_t\ba(\bx,t)]\psi(\by,t)
\eea

\section{Calculations of dual electric and magnetic fields}\label{sec:dual-calculation}

We would like to calculate the dual electromagnetic fields from the dual potential defined in Eq.(\ref{dual-A}). First let us calculate $\bar{\bB}\equiv \nabla\times\bar{\bA}$. It is found to be
\bea \bar{\bB}(\bx,t)  &=& -e\int d\by \bG(\bx,\by)\rho_e(\by,t) -g\int d\by \bG(\bx,\by)\times \bj_m (\by,t)  \nn \\ &&  +\partial_t\ba^T(\bx,t) \nn \\ &=&  e\nabla [\int d\by D(\bx,\by)\rho_e(\by,t)]  -\be'(\bx,t) +\partial_t\ba^T(\bx,t) \nn \\ &=&  \nabla a_0(\bx,t) -\be'(\bx,t) +\partial_t\ba^T(\bx,t) \nn \\ &=& -\bE(\bx,t),  \label{dual-B-2} \eea where $\ba_T$ is the transverse part of $\ba$. Here we have chosen the Coulomb gauge ($\lambda=0$ gauge), thus $e\int d\by D(\bx,\by)\rho_e(\by,t)=a_0(\bx,t)$.
In the derivation of Eq.(\ref{dual-B-2}), a useful intermediate step is to rewrite $ \int d\by \nabla_\by  D(\bx,\by) \times [ \nabla_\by\times\partial_t\ba(\by,t)]$ according to the well known  formula $\nabla(\bX\cdot\bY) = \bX\times(\nabla\times \bY)+ (\bX\cdot\nabla)\bY+ \bY\times(\nabla\times \bX)+ (\bY\cdot\nabla)\bX$, with $\bX=\partial_t \ba(\by,t)$ and $\bY=\nabla_\by D(\bx,\by)$\footnote{The calculation for Eq.(\ref{dual-B-2}) is easier if we do it in the momentum space and then translate back into spacetime.}.

Now let us turn to the dual electric field, namely, $\bar{\bE}\equiv -\nabla\bar{A}_0 -\partial_t\bar{\bA}$. According to Eq.(\ref{dual-A}), we find that  \bea \bar{\bE} (\bx,t)  =\bb'(\bx,t) - e\int d\by \bG(\bx,\by)\times\bj_e(\by,t) \nn \\ -\int d\by D(\bx,\by)\nabla_\by\times\partial_t^2 \ba^T(\by,t)  -g\int d\by D(\bx,\by)\partial_t\bj^T_m(\by,t)   \eea  Now it becomes more convenient to translate this equation into momentum space, wherein it reads \bea \bar{\bE}(q)  = \bb'(q) + e\frac{i\bq\times \bj_e(q)}{|\bq|^2}  + \frac{i q_0^2}{|\bq|^2}\bq\times\ba(q) + g\frac{iq_0}{|\bq|^2}\bj_m^T(q)   \eea in which $q=(q_0,\bq)$. Taking advantage of the transverse part of the quantum equations of motion associated with the Maxwell equations $(d)$, or more explicitly, Eq.(\ref{maxwell-d-1}), which in momentum space is given as \bea (q_0^2-|\bq|^2  )\ba^T -g q_0\frac{\bq\times\bj_m^T(q)}{|\bq|^2} + e\bj^T_e (q) =0 \,   , \eea we can obtain that \bea \bar{\bE}(q) = i\bq\times \ba^T(q) + \bb'(q), \eea which in spacetime is exactly \bea \bar{\bE}(\bx,t)&=&\nabla\times\ba^T(\bx,t)+ \bb'(\bx,t) \nn \\ &=& \bB(\bx,t)  \label{dual-E-2} \eea

\section{Comparison with the Hamiltonian formalism}\label{sec:equivalence}

The apparent differences between the present approach and
Tu-Wu-Yang's Hamiltonian approach\cite{tu1978}, other than that
ordinary numbers instead of noncommutative operators are used here,
appear in the form of monopole-photon interaction and the presence of
the last term in Eq.(\ref{final'}) or Eq.(\ref{final-coulomb}). In
spite of these differences in formulation, we shall show that our
approach is equivalent to the Tu-Wu-Yang approach\cite{tu1978}.

Let us start from the present Lagrangian formulation. We shall work in the Coulomb gauge (the $\lambda=0$ gauge), in which only transverse part of $\ba$ appears. From Eq.(\ref{monopole-dirac}) (or its promotion as an operator equation in the path integral approach), we know that the effective vector potential to which monopoles are coupled is \bea \bar{\bA}=\bar{\ba}' + \bar{\ba}^T, \eea in which \bea \bar{\ba}'  = -e\int d\by \bA(\bx,\by)\rho_e(\by,t), \label{coupling-1} \eea and \bea \bar{\ba}^T(\bx,t) =  \int
d\by D(\bx,\by) \nabla_\by\times\partial_t\ba^T(\by,t) \nn \\  + g\int d\by
D(\bx,\by)\bj_m^T(\by,t)  \label{bar-a-T} \eea Let us compare this with the gauge potential to which monopoles are coupled in Ref.\cite{tu1978}, namely the Eq.(3.2) in Ref.\cite{tu1978}.  Eq.(\ref{coupling-1}) can apparently be identified with the first term of Eq.(3.2) of Ref.\cite{tu1978}. Now let us investigate Eq.(\ref{bar-a-T}). According to the formulas $\bar{\bB}= \nabla\times\bar{\ba}^T+\bar{\bb}'$, $ \bE = -\partial_t \ba^T -\nabla {a}_0 +\be'$, and the dual relation $\bar{\bB}=-\bE$, we have \bea \partial_t\ba^T  =\nabla\times\bar{\ba}^T+ \be' \label{compare-1} \eea Eq.(\ref{compare-1}) can also be obtained by directly calculating the curl of $\bar{\ba}^T$. Similarly, we have \bea \partial_t\bar{\ba}^T  = -\nabla\times \ba^T + \bar{\be}' \label{compare-2} \eea  We emphasize that Eq.(\ref{compare-1}) and Eq.(\ref{compare-2}) can be regarded as operator equations, according to the correspondence between classical equations and operator equations in the path integral formalism\footnote{ See Sec.\ref{sec:motion} of the present paper or Ref.\cite{peskin1995} for the promotion of classical equations to operator equations in the path integral approach. }.

Now the connection to the Tu-Wu-Yang approach becomes clear.
In the Tu-Wu-Yang formalism\cite{tu1978}, the electron-photon and monopole-photon interaction are given by their Eq.(3.2) and Eq.(3.3), in which the evolution of $\bA^T$ and $\bB^T$ \footnote{Do not confuse about the notational differences between our paper and Ref.\cite{tu1978}. In Ref.\cite{tu1978} both $\bA^T$ and $\bB^T$ refer to gauge potential. In our paper $\bB$ refers to the magnetic field, while $\bB^T$ never appear. } is fully determined by their Eq.(4.6) and Eq.(4.7). It is evident that our Eq.(\ref{compare-1}) and Eq.(\ref{compare-2}) take the same forms as Eq.(4.6) and Eq.(4.7) in Ref.\cite{tu1978}, which allows us to identify $\ba^T$ and $\bar{\ba}^T$ in our paper as $\bA^T$ and $\bB^T$ in Ref.\cite{tu1978} respectively. Therefore, the couplings of monopoles to electromagnetic field are essentially the same in these two approaches, though they are seemingly different.

To be more explicit, we can solve Eq.(4.6) of Ref.\cite{tu1978}, and find that their $\bB^T$ can be expressed in terms of $\bA^T$ as \bea \bB^T(\bx,t)=\int d\by D(\bx,\by)\nabla_\by\times \partial_t \bA^T(\by,t)  \nn \\ + g\int d\by D(\bx,\by)\bj_m^T(\by,t), \eea
which takes the same form as our Eq.(\ref{bar-a-T}), therefore, we see again that $\bB^T(\bx,t)$ in Ref.\cite{tu1978} can be identified as $\bar{\ba}^T$ in our paper (the identification of $\bA^T$ in Ref.\cite{tu1978} as our $\ba^T$ is obvious). Therefore, the monopole-photon interaction turns out to be essentially the same in Ref.\cite{tu1978} and in our paper.

\bibliography{Monopole}

\end{document}